\documentclass[proof]{WileyASNA-v1}

\articletype{Original Article}
\received{TBD}\revised{TBD}\accepted{TBD}
\AtBeginDocument{%
}

\usepackage{hyperref}

\begin{document}

\title{Spectroscopic Studies of Meteors within the Skytinel Network}

\author[astrum,skytinel]{Ernest Świerczyński*}
\author[skytinel,ptmet]{Mateusz Żmija}
\author[skytinel]{Paweł Zaręba}
\author[skytinel]{Tomasz Żywczak}

\authormark{ŚWIERCZYŃSKI \textsc{et al.}}

\address[astrum]{%
\orgname{Astrum Narusa Foundation},
\orgaddress{\country{Poland}}}

\address[skytinel]{%
\orgname{Skytinel},
\orgaddress{\country{Poland}}}

\address[ptmet]{%
\orgname{Polish Meteoritical Society},
\orgaddress{\country{Poland}}}

\corres{*Ernest Świerczyński. \email{eswierczynski@narusa.pl}}

\abstract{We present 12 low-resolution meteor spectra recorded at four
stations of the Skytinel network in Poland, focusing on consistent
reduction and comparison of Fe, Mg, and Na emission features.
Spectra spanning approximately 350--850 nm were extracted,
background-subtracted, and wavelength-calibrated using IRAF.
Corrections accounted for the wavelength-dependent quantum
efficiencies of Sony IMX249 and IMX273 detectors and atmospheric
attenuation as a function of observing geometry.
The corrected spectra were decomposed into Fe, Mg, and Na
components using selected atomic transitions, temperature-dependent
emissivities, and pseudo-Voigt line profiles.
Integrated Fe~I-15, Mg~I-2, and Na~I-1 intensities were compared
in a ternary diagram. Most spectra overlap with or lie near the
normal-meteor distribution in reference samples, while some
extend toward Fe-rich or Na-rich regions. The most Fe-dominated
event assigns approximately 81\% of the summed emission to
Fe~I-15. These ratios characterize relative emission, not
elemental abundances or bulk composition. Combined with
trajectory and photometric data, spectral classification may
help constrain plausible material properties in dark-flight
models and refine predicted meteorite search areas.}

\keywords{meteors, meteor spectroscopy, meteoroids, Skytinel, Fe--Mg--Na classification, atmospheric extinction}
\maketitle

\begin{table*}[b]
\caption{Skytinel stations providing the meteor spectra analysed in this work. Geographic coordinates are given in decimal degrees, and altitude refers to the station elevation above sea level.}
\label{tab:stations}
\centering
\begin{tabular*}{500pt}{@{\extracolsep\fill}lcccccc@{\extracolsep\fill}}
\toprule
Station & Location & Altitude (m) & Latitude ($^\circ$N) & Longitude ($^\circ$E) & Sensor & Spectral frame (px) \\
\midrule
SN01 & Stanisław Dolny    & 300 & 49.89 & 19.64 & IMX249 & 1600x1200 \\
SN38 & Chrzanów Mały      & 100 & 52.13 & 20.63 & IMX249 & 1920x1200 \\
SN47 & Zabierzów Bocheński & 200 & 50.07 & 20.31 & IMX249 & 1920x1200 \\
SN48 & Frombork/Narusa    & 20  & 54.34 & 19.66 & IMX273 & 1440 x 1080 \\
\bottomrule
\end{tabular*}
\end{table*}

\section{Introduction}\label{sec:intro}

Meteor spectroscopy provides information on the radiating material produced during meteoroid ablation. In low-resolution spectra, emission from neutral and ionized metals forms characteristic features, with Fe, Mg, and Na among the most commonly observed species, although their relative strengths vary considerably and individual components may be weak or absent. These differences have long been used as a practical description of meteor spectra and as the basis of spectral classification \shortcites{Borovicka2005}\citet{Borovicka2005}. This approach was further developed in the catalogue of representative meteor spectra by \shortcites{Vojacek2015}\citet{Vojacek2015}, where spectral properties were considered together with meteor velocities and dynamical information to characterize different meteoroid populations.

An extension to brighter and generally larger meteoroids was provided by \shortcites{Matlovic2019}\citet{Matlovic2019}. Their study identified a Fe-rich spectral group located between the normal and iron classes in the Fe--Mg--Na representation, in a region sparsely populated in earlier samples of fainter meteors. This provides a particularly useful comparison for the bright meteor spectra considered here.

The Skytinel network is a distributed meteor-observation project operating stations in Poland and neighbouring regions. Its development, observing equipment, software, and trajectory and orbital calculations have been described by \citet{Zmija2025}. Skytinel focuses particularly on events that may result in meteorite falls and their subsequent recovery; during its first two years of operation, the network contributed to the recovery of three meteorites \shortcites{Bischoff2026Drelow,Zmija2026Iron}\citep{Bischoff2026Drelow,Zmija2026Iron}. In this context, spectroscopic discrimination between broadly stony and iron-rich material may provide an additional constraint on the range of plausible bulk densities adopted in dark-flight calculations and thus help to reduce the predicted meteorite search area.

The present paper examines low-resolution spectra obtained at four
Skytinel stations. Relative Fe, Mg, and Na emission intensities
are derived using a simplified three-component spectral model
based on selected atomic transitions and compared using a ternary
Fe--Mg--Na representation. Particular attention is given to the
spectral diversity of bright meteors and their comparison with
established meteor spectral classes. The resulting intensities
are treated as relative emission measures rather than absolute
elemental abundances. The broader aim is to assess how such
spectroscopic information can complement the analysis of
potential meteorite falls.

\section{Observations and instrumentation}\label{sec:obs}
\subsection{Spectroscopic stations}
Continuous spectroscopic observations are currently performed at four Skytinel stations: SN01 at Stanisław Dolny, SN38 at Chrzanów Mały, SN47 at Zabierzów Bocheński, and SN48 at the Narusa Observatory near Frombork. The station codes are used throughout the paper to monitor possible systematic differences between individual stations at each stage of the analysis, including those related to their different locations and camera models (Table~\ref{tab:stations}). The spectrographs use a 6-mm lens and diffraction gratings of 1000 lines mm$^{-1}$, but SN01, SN38, and SN47 use cameras based on the Sony IMX249 sensor, whereas SN48 uses a Sony IMX273 sensor.

\begin{figure*}[t]
\centerline{\includegraphics[width=0.7\textwidth]{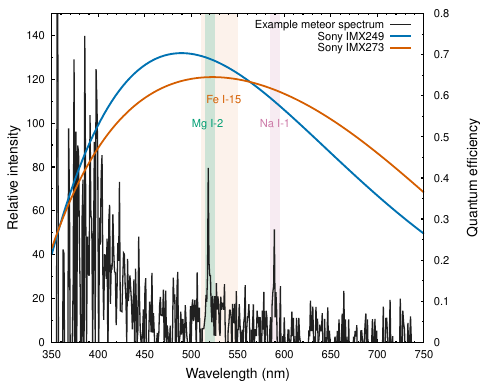}}
\caption{Quantum-efficiency curves of the Sony IMX249 and IMX273
sensors compared with a representative meteor spectrum. The shaded
regions indicate the principal spectral regions containing the
Fe~I-15, Mg~I-2, and Na~I-1 features used in the analysis. These
features fall within the high-sensitivity range of both detectors.}\label{fig:qe}
\end{figure*}

The two sensors have different wavelength-dependent quantum efficiencies.
The spectral-response curves used for the detector correction were taken
from the technical documentation provided by the camera manufacturer
\cite{Daheng2024}. This difference in detector response is relevant when
spectra obtained at different stations are compared. Figure~\ref{fig:qe}
shows the response curves adopted in the analysis. The corresponding
detector correction was applied separately to spectra recorded with the
IMX249 and IMX273 sensors. This correction accounts for the wavelength dependence of the detector
response and should not be interpreted as a full spectrophotometric
calibration of the complete optical system, which would additionally
require the wavelength-dependent response of the lens and diffraction
grating.

The effective spectral resolution, estimated from the mean FWHM of
isolated emission features in the calibrated meteor spectra, was
approximately 1.2~nm, corresponding to a resolving power of
$R \sim 433$ in the Fe--Mg spectral region.

\begin{table}[t]
\caption{Meteor spectra analysed in this work. Times are given in local time.
The apparent elevation of each meteor was determined astrometrically from
the spectral frame.}
\label{tab:spectral_sample}
\centering
\begin{tabular}{llccccc}
\toprule
ID & Station & Date & Time & Elev. ($^\circ$) \\
\midrule
SN01\_240813 & SN01 & 2024-08-13 & 23:40 & 34 \\
SN01\_250105 & SN01 & 2025-01-05 & 03:45 & 19 \\
SN01\_250120 & SN01 & 2025-01-20 & 03:13 & 12 \\
SN38\_260404 & SN38 & 2026-04-04 & 01:59 & 35 \\
SN38\_260912 & SN38 & 2026-09-12 & 22:53 & 30 \\
SN47\_260521 & SN47 & 2026-05-21 & 01:52 & 10 \\
SN48\_260103 & SN48 & 2026-01-03 & 23:08 & 15 \\
SN48\_260216 & SN48 & 2026-02-16 & 04:13 & 13 \\
SN48\_260521 & SN48 & 2026-05-21 & 00:18 & 18 \\
SN48\_260603 & SN48 & 2026-06-03 & 23:52 & 24 \\
SN48\_260626 & SN48 & 2026-06-26 & 01:52 & 38 \\
SN48\_260731 & SN48 & 2026-07-31 & 03:57 & 26 \\
\bottomrule
\end{tabular}
\end{table}

\subsection{Spectral sample}

The material analysed here consists of low-resolution meteor spectra recorded between 2024 and 2026 during routine operation of the Skytinel network. The individual observations, including station, date, time, apparent elevation, and spectrum identifier, are summarized in Table~\ref{tab:spectral_sample}. The sample includes events for which the Fe, Mg, and Na spectral regions could be measured and compared using the procedure described below.

In addition to the first-order spectrum, the zero-order image of the meteor was recorded in the same frame against the stellar background. Identification of field stars and astrometric calibration allowed the apparent elevation of the meteor above the horizon to be determined directly from the spectral frame. This measured elevation provides the observational geometry required for the atmospheric correction described in Section~\ref{sec:atm}.

\section{Spectral reduction}\label{sec:red}

Spectral extraction and calibration were performed using tools from the Image Reduction and Analysis Facility (IRAF). The processing included subtraction of the local background, extraction of the meteor spectrum, and wavelength calibration. These are standard operations in IRAF spectral reduction, in which background regions adjacent to the extracted spectrum can be used to estimate and remove the local background before the one-dimensional spectrum is produced \citep{Tody1986,Tody1993}.

\begin{table}[b]
\caption{Principal spectral features used, where identifiable, as reference features for wavelength calibration.}
\label{tab:calibration_lines}
\centering
\begin{tabular}{lcc}
\toprule
Element & Identification & Wavelength (nm) \\
\midrule
Ca II & H and K doublet       & 393.37,\ 396.85 \\
Ca I  &                       & 422.67 \\
Mg II & multiplet             & 448.12 \\
Mg I  & triplet               & 518.2 \\
Na I  & D doublet             & 589.2 \\
Si II   &                       & 636.0 \\
O I   & multiplet             & 777.4 \\
N I   & multiplet             & 821.6 \\
O I   & multiplet             & 844.6 \\
\bottomrule
\end{tabular}
\end{table}

The spectra cover approximately 350--850 nm. Wavelength calibration was based on identified meteor and atmospheric emission features distributed across the spectral range. The principal reference features are listed in Table~\ref{tab:calibration_lines}.

\begin{figure*}[b]
\centerline{\includegraphics[width=1.0\textwidth]{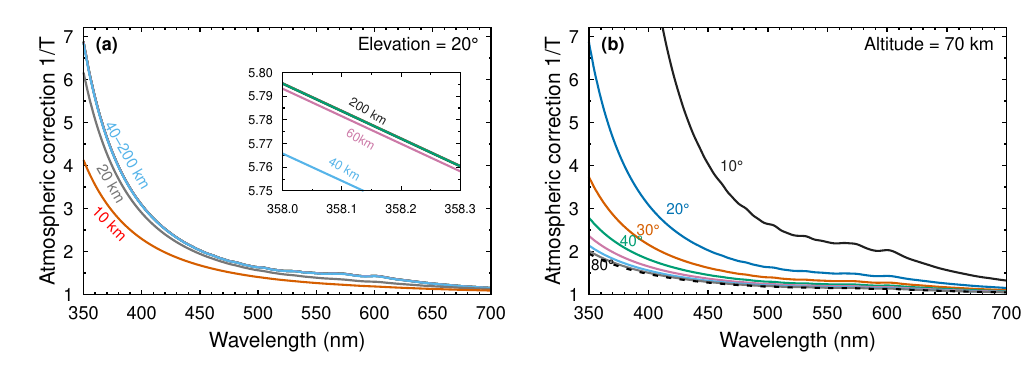}}
\caption{Dependence of the atmospheric correction function \(1/T(\lambda)\) on the assumed meteor geometry. (a) Correction curves calculated for meteor altitudes between 20 and 200 km at a fixed apparent elevation of \(20^\circ\). Above approximately 40 km, the curves converge rapidly and the differences become small, as illustrated by the inset. (b) Correction curves calculated for a fixed meteor altitude of 70 km and apparent elevations between \(10^\circ\) and \(80^\circ\). In contrast to source altitude, changes in apparent elevation produce clearly distinguishable correction curves throughout the investigated range. This is advantageous for the present analysis because the apparent elevation can be determined astrometrically from the spectral frame, independently of a complete three-dimensional trajectory solution.\label{fig:atm}}
\end{figure*}

The measured spectrum can be written schematically as
\begin{equation}
S_{\rm obs}(\lambda) \propto S_{\rm in}(\lambda)Q(\lambda),
\end{equation}
where $Q(\lambda)$ represents the wavelength-dependent response of the detector. The recorded spectra were therefore corrected using the response curve appropriate to the sensor used for the observation. IMX249 response data were applied to SN01, SN38, and SN47 spectra, and IMX273 response data to SN48 spectra. This correction precedes the comparison of Fe, Mg, and Na intensities between stations.

\section{Atmospheric transmission}\label{sec:atm}

The radiation emitted by a meteor is attenuated between the emitting point and the observer. The amount of attenuation depends on wavelength and on the atmospheric column crossed by the radiation. Consequently, both the physical meteor height and its elevation above the horizon are relevant to the correction.

\subsection{Reference atmosphere}

The vertical atmospheric profile was generated from the U.S. Standard Atmosphere 1976 \citep{USSA1976}. The numerical implementation uses a grid from 0 to 100 km with a spacing of 100 m. At each altitude the model supplies the quantities required to describe the density and composition of the atmosphere. For the present calculation, the relevant output is the altitude dependence of the number densities of the principal atmospheric species together with temperature, pressure, and density. The purpose is not to reproduce the instantaneous meteorological state at each station, but to use one physically defined reference atmosphere consistently for all spectra.

\subsection{Rayleigh scattering and molecular absorption}

The total extinction coefficient was separated into Rayleigh scattering, ozone absorption, and a simplified molecular oxygen contribution. The total vertical optical depth to a meteor at height $h$ is
\begin{equation}
\tau_{\rm v}(\lambda,h)=\int_0^h
[\alpha_{\rm R}(\lambda,z)+\alpha_{\rm O_3}(\lambda,z)+\alpha_{\rm O_2}(\lambda,z)]\,dz.
\end{equation}

For Rayleigh scattering, effective cross-sections at 550 nm are used for N$_2$, O$_2$, Ar, O, He, and H and scaled as $\lambda^{-4}$,
\begin{equation}
\alpha_{\rm R}(\lambda,z)=\sum_i n_i(z)\sigma_i(550\,{\rm nm})
\left(\frac{550\,{\rm nm}}{\lambda}\right)^4.
\end{equation}
Thus the short-wavelength part of a meteor spectrum is attenuated more strongly than the longer-wavelength part (Figure~\ref{fig:atm}).

Ozone absorption is calculated from the ozone number density and wavelength-dependent absorption cross-section,
\begin{equation}
\alpha_{\rm O_3}(\lambda,z)=n_{\rm O_3}(z)\sigma_{\rm O_3}(\lambda).
\end{equation}
The cross-section data used here are based on the laboratory measurements of \shortcites{Gorshelev2014}\citet{Gorshelev2014} and \shortcites{Serdyuchenko2014}\citet{Serdyuchenko2014}, which provide high-resolution ozone absorption cross-sections over 213--1100 nm and at a range of temperatures. These data are used because ozone produces a strongly wavelength-dependent contribution to atmospheric transmission, particularly toward the ultraviolet and blue part of the spectral range.

A simplified O$_2$ term is included around the atmospheric A band near 760 nm. In the numerical model it is represented only as an approximate absorption contribution in a narrow wavelength interval. It is not a line-by-line model of molecular oxygen and is not used to interpret individual O$_2$ lines.

\subsection{Meteor--observer path}

The vertical optical depth is converted to the slant path using a plane-parallel approximation. For zenith angle $\zeta$,
\begin{equation}
\tau_{\rm s}(\lambda,h,\zeta)=\frac{\tau_{\rm v}(\lambda,h)}{\cos\zeta},
\end{equation}
with the atmospheric transmission
\begin{equation}
T(\lambda,h,\zeta)=\exp[-\tau_{\rm s}(\lambda,h,\zeta)].
\end{equation}

When an independently determined meteor altitude was not available,
a representative altitude of 70~km was adopted for the atmospheric
correction. As shown in Figure~\ref{fig:atm}, the correction
depends only weakly on the assumed source altitude above approximately
40~km, whereas the apparent elevation has a substantially stronger
effect. This is consistent with luminous meteor trajectories generally occurring well above the dense lower atmosphere \citep{Ceplecha1998}.

\begin{table}[!tbp]
\centering
\caption{Atomic parameters adopted in the three-component
spectral emission model. Wavelengths $\lambda$, upper-level
excitation energies $E_u$, and logarithmic weighted
oscillator strengths $\log(gf)$ characterize the individual
transitions. Horizontal lines separate the Fe, Mg, and Na
components, scaled independently by $C_{\rm Fe}$,
$C_{\rm Mg}$, and $C_{\rm Na}$, respectively.}
\label{tab:atomic_lines}

\begin{tabular}{lrrr}
\hline
Multiplet &
$\lambda$ (nm) &
$E_u$ (eV) &
$\log(gf)$ \\
\hline

Fe I-37
 & 516.7490 & 3.88431 & -1.160 \\
Fe I-15
 & 526.9540 & 3.21085 & -1.321 \\
 & 532.8040 & 3.24201 & -1.466 \\
 & 537.1490 & 3.26619 & -1.645 \\
 & 539.7130 & 3.21222 & -1.993 \\
 & 540.5770 & 3.28355 & -1.844 \\
 & 542.9700 & 3.24144 & -1.879 \\
 & 543.4520 & 3.29242 & -2.122 \\
 & 544.6920 & 3.26623 & -1.914 \\
 & 545.5610 & 3.28360 & -2.091 \\

\hline

Mg I-2
 & 516.7322 & 5.10783 & -0.870 \\
 & 517.2684 & 5.10783 & -0.393 \\
 & 518.3604 & 5.10783 & -0.167 \\

\hline

Na I-1
 & 588.9950 & 2.10501 &  0.108 \\
 & 589.5920 & 2.10288 & -0.194 \\

\hline
\end{tabular}
\end{table}

\section{Spectral Model}

\begin{figure*}[t]
\centerline{\includegraphics[width=0.84\textwidth]{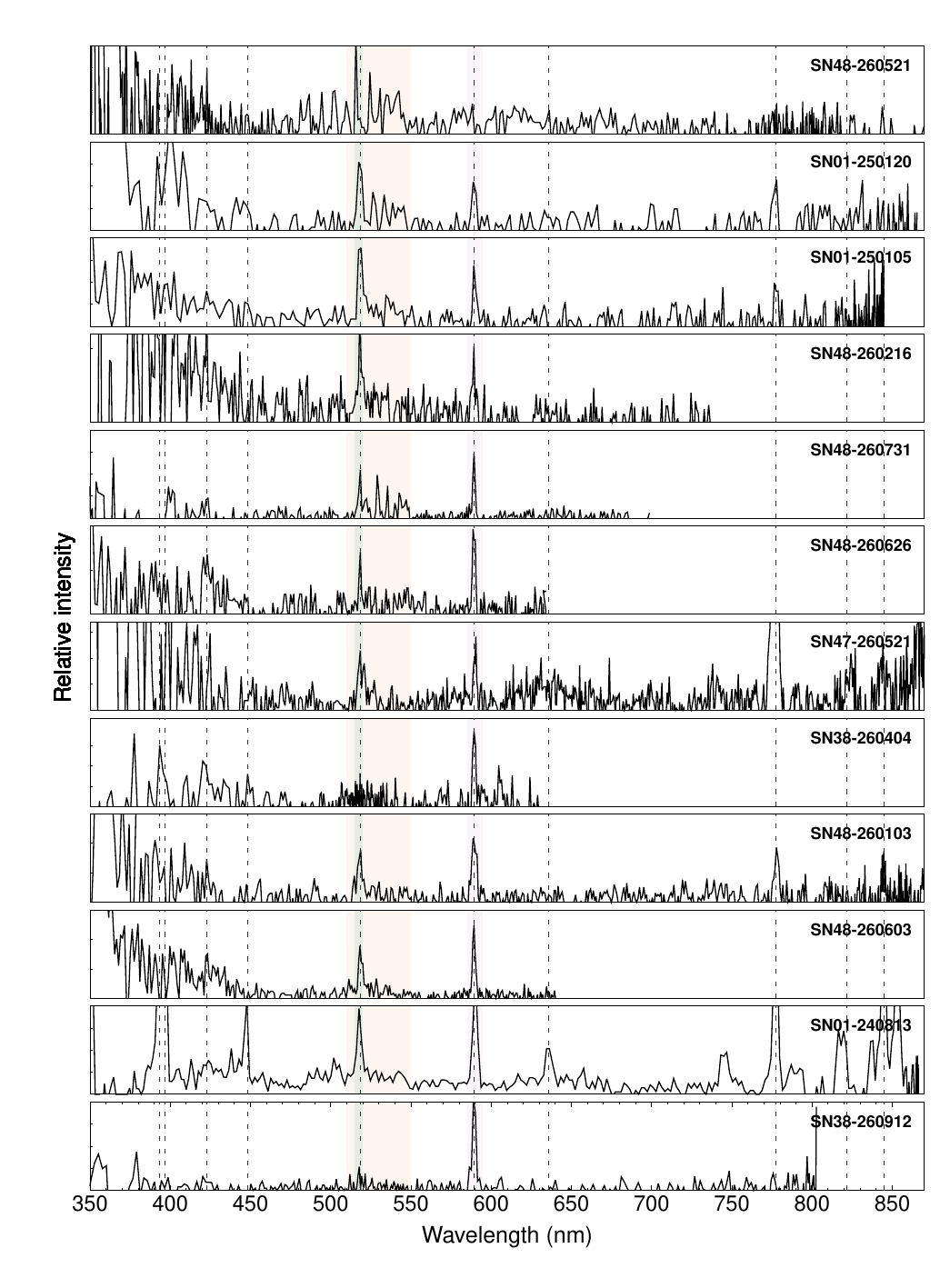}}
\caption{Corrected meteor spectra.
Vertical dashed lines mark selected spectral features from
Table~\ref{tab:calibration_lines}, while the shaded areas
indicate the principal spectral regions containing the
Fe~I-15, Mg~I-2, and Na~I-1 features. }
\label{fig:sequence}
\end{figure*}

The spectra differ substantially in signal-to-noise ratio and
in the amount of small-scale structure around the principal
features. The relative intensities of Mg~I-2 and neighbouring
Fe emission, as well as more isolated features such as Na~I-1,
vary significantly between spectra.
The spectral diversity illustrated in Figure~\ref{fig:sequence}
highlights the challenges of quantitative comparison and
motivates the use of a simplified spectral model. The analysis focuses on the most prominent Fe~I-15, Mg~I-2,
and Na~I-1 multiplets, which form the basis of the widely
used Fe--Mg--Na comparison of meteor spectra
\citep{Borovicka2005}.

The relative prominence of these
features varies considerably among the corrected spectra
shown in Figure~\ref{fig:sequence}. Some spectra show a
pronounced Fe-dominated structure in the 527--546~nm
region, whereas in others the Na~I-1 feature near 589~nm
is one of the strongest features in the spectrum. The Mg~I-2 feature near 517~nm also varies substantially
and, at the present spectral resolution, is blended with
nearby Fe emission. The Fe and Mg contributions were
therefore modelled simultaneously rather than assigning
the complete 517-nm feature to Mg.

The selected Fe~I, Mg~I, and Na~I transitions listed in
Table~\ref{tab:atomic_lines} are based on multiplets commonly
used in meteor spectral modelling and classification
\citep{Borovicka2005,Vojacek2015} and correspond to prominent
features in the observed spectra. These transitions were
adopted to construct a simplified three-component spectral
model intended to provide a consistent decomposition of the
Fe, Mg, and Na emissions for integration and comparison
between meteors, rather than to reproduce every local
spectral feature.

We assumed that individual emission features were represented
by pseudo-Voigt profiles, using the Thompson--Cox--Hastings
approximation \shortcites{Thompson1987}\citep{Thompson1987}
to the Voigt line profile \shortcites{Armstrong1967}
\citep{Armstrong1967}. Their application to synthetic meteor
spectra is described, for example, by \shortcites{Segon2024}\citet{Segon2024}.

For a transition centred at $\lambda_i$, the line profile
is denoted by

\begin{equation}
P_i(\lambda) =
V(\lambda;\lambda_i,\sigma_{\mathrm{line}},\gamma),
\end{equation}

where $\sigma_{\mathrm{line}}$ and $\gamma$ are the Gaussian
and Lorentzian widths, respectively. The Gaussian width
was represented as

\begin{equation}
\sigma_{\mathrm{line}} =
\sqrt{\sigma_0^2+\sigma_{\mathrm{inst}}^2},
\end{equation}

with $\sigma_{\mathrm{inst}}$ adjusted to the effective
spectral resolution of each instrumental configuration.

\begin{figure*}[t]
\centering
\includegraphics[width=0.9\textwidth]{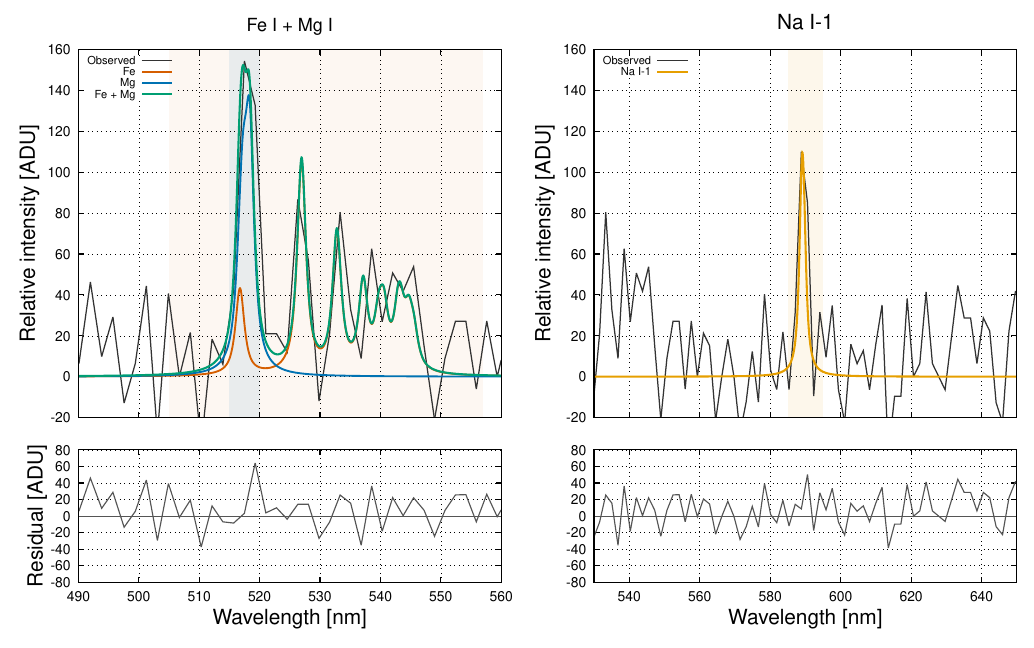}
\caption{Example of the three-component spectral model fitted to
an observed meteor spectrum. Panel (a) shows the Fe and Mg
components and their combined model in the 490--560~nm
region, while panel (b) presents the Na component in the
region of the Na~I-1 doublet. The observed spectra are
shown together with the corresponding modelled emission.
Shaded areas indicate the principal spectral regions
containing the Fe~I-15, Mg~I-2, and Na~I-1 features
and do not represent the integration limits.
Vertical dashed lines mark the selected Fe~I-15 and
Na~I-1 transitions. The lower panels show the residuals,
defined as observed minus model intensity.}
\label{fig:spectral_fit}
\end{figure*}

\newpage
\phantom{aaa}
\newpage
\phantom{aaa}
\newpage

The relative line emissivities were calculated assuming
optically thin emission and Boltzmann-distributed populations
of excited atomic levels. The oscillator strengths were
converted into weighted transition probabilities, which,
together with the upper-level excitation energies, determine
the temperature-dependent emissivity function
$\Phi_i(T)$, defined as

\begin{equation}
\Phi_i(T) =
g_{u,i}A_{ul,i}
\exp\left(-\frac{E_{u,i}}{k_B T}\right)
\frac{hc}{\lambda_i}.
\end{equation}

Here, $g_{u,i}$ is the statistical weight of the upper level.
Atomic wavelengths, upper-level excitation energies, and
oscillator strengths were adopted from the NIST Atomic
Spectra Database \shortcites{NIST_ASD}\citep{NIST_ASD}. The transitions
and their parameters used in the model are listed in
Table~\ref{tab:atomic_lines}.

\begin{table*}[b]
\centering
\caption{Integrated intensities of the modelled Fe~I-15,
Mg~I-2, and Na~I-1 emission components and their normalized
contributions to the Fe--Mg--Na ternary representation.}
\label{tab:integrated_components}

\begin{tabular*}{\textwidth}{@{\extracolsep{\fill}}lrrrrrr@{}}
\hline
Spectrum ID
& $I_{\mathrm{Fe}}$
& $I_{\mathrm{Mg}}$
& $I_{\mathrm{Na}}$
& Fe (\%)
& Mg (\%)
& Na (\%) \\
\hline
SN48\_260521 & 1194.00 & 122.66 & 191.63 & 79.2 & 8.1 & 12.7 \\
SN01\_250120 &  876.70 & 505.93 &  350.01 & 50.6 & 29.2 & 20.2 \\
SN48\_260216 &  338.16 & 303.56 &  150.00 & 42.7 & 38.3 & 18.9 \\
SN48\_260731 &  868.24 & 247.44 & 1185.53 & 37.7 & 10.8 & 51.5 \\
SN01\_250105 &  596.24 & 622.00 &  379.80 & 37.3 & 38.9 & 23.8 \\
SN48\_260103 &  375.73 & 303.56 &  400.01 & 34.8 & 28.1 & 37.1 \\
SN47\_260521 &  438.35 & 379.44 &  450.01 & 34.6 & 29.9 & 35.5 \\
SN48\_260626 &  252.07 & 243.65 &  296.04 & 31.8 & 30.8 & 37.4 \\
SN48\_260603 &  290.39 & 320.06 &  345.94 & 30.4 & 33.5 & 36.2 \\
SN38\_260404 &  375.73 & 303.56 &  600.02 & 29.4 & 23.7 & 46.9 \\
SN38\_260912 &  238.50 & 178.31 &  617.17 & 23.1 & 17.2 & 59.7 \\
SN01\_240813 &  310.16 & 644.28 &  745.00 & 18.3 & 37.9 & 43.8 \\
\end{tabular*}

\end{table*}

The weighted transition probabilities were calculated as

\begin{equation}
g_{u,i}A_{ul,i} =
\frac{6.670\times10^{15}}{\lambda_i^2}
10^{\log(gf)_i},
\end{equation}

where $\lambda_i$ is expressed in \AA.

Using the previously defined line profile $P_i(\lambda)$,
the spectral contribution of an individual transition is

\begin{equation}
F_i(\lambda,T) = \Phi_i(T)P_i(\lambda).
\end{equation}

The contribution of each elemental component was obtained
by summing its selected transitions:

\begin{equation}
F_X(\lambda,T) =
\sum_{i\in X}F_i(\lambda,T),
\qquad X\in\{\mathrm{Fe,Mg,Na}\}.
\end{equation}

Finally, the complete spectral model was constructed as
the sum of three independently scaled components:

\begin{equation}
F_{\mathrm{model}}(\lambda,T) =
C_{\mathrm{Fe}}F_{\mathrm{Fe}}(\lambda,T)
+ C_{\mathrm{Mg}}F_{\mathrm{Mg}}(\lambda,T)
+ C_{\mathrm{Na}}F_{\mathrm{Na}}(\lambda,T),
\end{equation}

where $C_{\mathrm{Fe}}$, $C_{\mathrm{Mg}}$, and
$C_{\mathrm{Na}}$ are scaling factors adjusted to reproduce
the observed spectral features.

The $F_{\mathrm{Fe}}$ component contains the Fe~I-37
transition blended with the Mg~I-2 triplet near 517~nm,
as well as the selected Fe~I-15 transitions listed in
Table~\ref{tab:atomic_lines}. However, only Fe~I-15
was included in the integrated Fe intensity used for
ternary classification.

The scaling coefficients $C_{\mathrm{Fe}}$, $C_{\mathrm{Mg}}$,
and $C_{\mathrm{Na}}$ were adjusted manually for each
observed spectrum to reproduce the relative strengths
of the selected emission features. This approach follows
the practice of visually matching modelled and observed
meteor spectra, as employed in earlier spectroscopic
studies \citep{Vojacek2015}.
An example of the resulting model applied to an observed
meteor spectrum is shown in Figure~\ref{fig:spectral_fit}.

\begin{figure}[t]
\centering
\includegraphics[width=0.9\columnwidth]{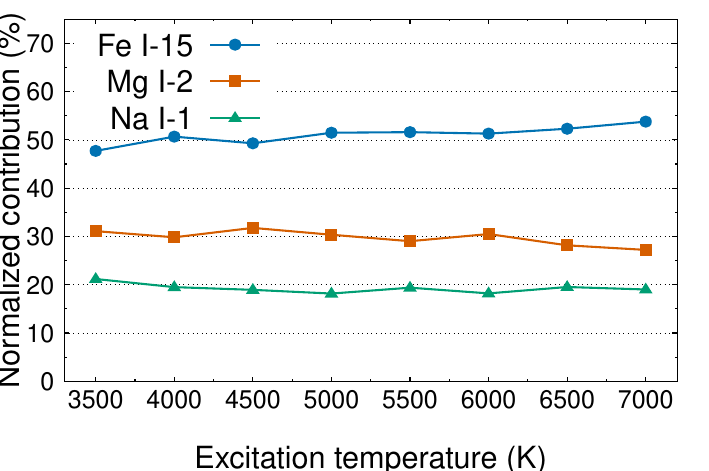}
\caption{Sensitivity of the normalized Fe~I-15, Mg~I-2, and
Na~I-1 contributions to the adopted excitation temperature
for a representative meteor spectrum. }
\label{fig:temperature_sensitivity}
\end{figure}

To evaluate the sensitivity of the derived Fe--Mg--Na
ratios to the adopted excitation temperature, the modelling
procedure was repeated for a representative spectrum over
the range $T=3500$--$7000$~K in steps of 500~K.
At each temperature, the scaling factors $C_{\mathrm{Fe}}$,
$C_{\mathrm{Mg}}$, and $C_{\mathrm{Na}}$ were readjusted
manually to obtain the best visual representation of the
observed spectral features. The resulting intensities of
Fe~I-15, Mg~I-2, and Na~I-1 were integrated and normalized
in the same manner as for the Fe--Mg--Na ternary diagram. The resulting temperature dependence of the normalized
component contributions is shown in
Figure~\ref{fig:temperature_sensitivity}.

\begin{figure*}[b]
\centerline{\includegraphics[width=0.9\textwidth]{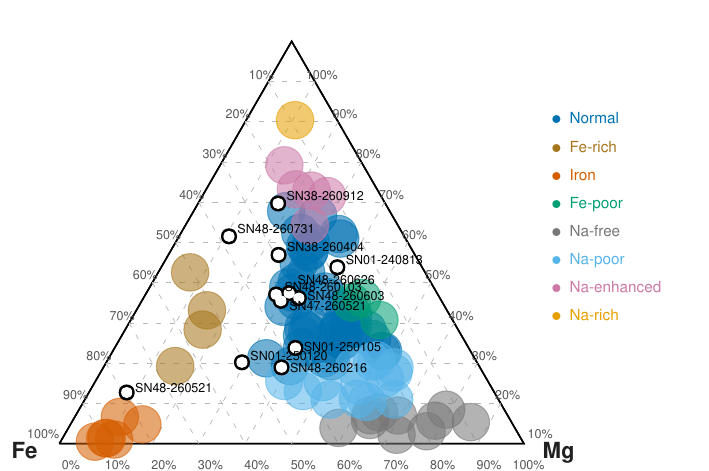}}
\caption{Ternary diagram of the normalized Fe, Mg, and Na integrated
emission components. Open black circles denote the Skytinel spectra.
Coloured translucent markers show the distributions of spectral
classes in the representative meteor catalogue of
\citet{Vojacek2015}; the Fe-rich class introduced by
\citet{Matlovic2019} is shown separately. The coloured regions
represent the distributions of the reference meteors and are not
intended as formal classification boundaries.
\label{fig:ternary}}
\end{figure*}


The normalized contributions of Fe I-15, Mg I-2, and Na I-1
remain relatively stable throughout the investigated
temperature range of 3500--7000 K. The Fe I-15 contribution
varies between approximately 48\% and 54\%, Mg I-2 between
27\% and 32\%, and Na I-1 between 18\% and 21\%.
No systematic redistribution of emission between the
three components is observed that would substantially
alter their relative contributions. These results indicate that, for the representative
spectrum considered here, the derived Fe--Mg--Na ratios
are only moderately sensitive to the adopted excitation
temperature when the component scaling coefficients
are readjusted at each temperature. The adopted range is consistent with the excitation
temperatures used by \citet{Borovicka2005} in theoretical
calculations of the relative intensities of the Fe~I-15,
Mg~I-2, and Na~I-1 multiplets for meteor spectral
classification. This sensitivity test assesses the
robustness of the derived component ratios and does not
constitute an independent determination of the excitation
temperature.

\section{Integrated intensities}

The quantities used for the Fe–Mg–Na comparison were obtained from the Fe I-15, Mg I-2, and Na I-1 model components. The total
area of each component was evaluated numerically over the common
wavelength interval 450--650~nm,

\begin{eqnarray}
I_{\rm Fe} &=C_{\mathrm{Fe}}& \int_{450}^{650}
F_{\rm Fe\,I-15}(\lambda)\,d\lambda\\
\end{eqnarray}

\begin{eqnarray}
I_{\rm Mg} &=C_{\mathrm{Mg}}& \int_{450}^{650}
F_{\rm Mg\,I-2}(\lambda)\,d\lambda,\\
I_{\rm Na} &=C_{\mathrm{Na}}& \int_{450}^{650}
F_{\rm Na\,I-1}(\lambda)\,d\lambda.
\end{eqnarray}

Because the integration was performed on the individual model components rather than directly on the observed spectrum, the common broad interval effectively includes the complete line profiles without adding neighbouring spectral features.
Additional transitions included to improve the representation of the observed spectrum were excluded from these integrals.

The resulting values are listed in
Table~\ref{tab:integrated_components}. They represent relative spectral
intensities and are not interpreted as direct measurements of elemental
abundances.

\section{Ternary Fe--Mg--Na representation}\label{sec:ternary}

For comparison between individual spectra, the three integrated intensities were normalized by their sum,
\begin{equation}
f_{\rm Fe}=\frac{I_{\rm Fe}}{I_{\rm Fe}+I_{\rm Mg}+I_{\rm Na}},
\end{equation}
with analogous expressions for $f_{\rm Mg}$ and $f_{\rm Na}$. The fractions satisfy
\begin{equation}
f_{\rm Fe}+f_{\rm Mg}+f_{\rm Na}=1.
\end{equation}

They were transformed to Cartesian coordinates for an equilateral ternary diagram using
\begin{equation}
X=f_{\rm Mg}+\frac{1}{2}f_{\rm Na},\qquad
Y=\frac{\sqrt{3}}{2}f_{\rm Na}.
\end{equation}

The comparison of relative Fe, Mg, and Na intensities follows the meteor spectral classification introduced by \citet{Borovicka2005}. The ternary diagram is used here in the same descriptive sense, representing relative emission intensities rather than bulk elemental fractions. The resulting distribution of the analysed spectra is shown in Figure~\ref{fig:ternary}.

\section{Results}\label{sec:results}

The analysed spectra occupy a broad range of positions in the ternary
diagram, demonstrating substantial differences in the relative
Fe~I-15, Mg~I-2, and Na~I-1 emission. Most of the Skytinel spectra
overlap with or lie close to the distribution of normal meteors in
\citet{Vojacek2015}, while individual events extend toward the
Fe-rich and Na-rich regions. This diversity is also apparent directly
in the corrected spectra shown in Figure~\ref{fig:sequence}.

SN48\_260521 represents the most strongly Fe-dominated
spectrum in the sample, with approximately 79\% of the
summed Fe--Mg--Na emission assigned to Fe~I-15,
compared with 8\% Mg~I-2 and 13\% Na~I-1.
In contrast, SN48\_260731 exhibits a relatively strong
Na~I-1 contribution (52\%) and a weak Mg~I-2
component (11\%). The highest relative Na~I-1
contribution is found in SN38\_260912, where sodium
accounts for approximately 60\% of the three-component
sum.

Comparison with the reference distributions shows that the majority
of the sample is compatible with the broad normal-meteor population,
rather than defining a distinct Skytinel group. The Fe-dominated tail
of the distribution nevertheless extends toward the Fe-rich region
identified by \citet{Matlovic2019}, which is sparsely populated in the
representative catalogue of \citet{Vojacek2015}. The present sample is
too small to establish whether this reflects a population-level
difference. It is, however, consistent with the occurrence of Fe-rich
spectra among the brighter meteors investigated by
\citet{Matlovic2019}.

\section{Discussion}\label{sec:discussion}

The Fe--Mg--Na distribution derived from the Skytinel spectra
is broadly consistent with the established diversity of meteor
spectral classes. Most events lie within or close to the
normal-meteor distribution of \citet{Vojacek2015}, while
individual spectra extend toward the Fe-rich and Na-rich
regions. The Fe-rich tail is consistent with the occurrence
of similar spectra among brighter meteors reported by
\citet{Matlovic2019}. However, differences in the methods
used to derive relative emission intensities, together with
the small and observationally selected Skytinel sample,
limit the quantitative interpretation of this comparison.

An important limitation is imposed by the spectral resolution of the
Skytinel instruments. The analysed spectra were individually checked
for saturation and rapid brightness variations that could
significantly affect the measured line ratios, and no such effects
were found to dominate the selected spectral frames. Residual background, wavelength-calibration
uncertainties, and the finite instrumental resolution may additionally
affect the component amplitudes. These effects are expected to be most important for weak components and for strongly blended features.

An additional source of systematic uncertainty is the
wavelength-dependent response of the optical system.
Although the detector quantum efficiency was corrected,
the spectral responses of the lens and diffraction grating
were not independently calibrated. Their combined effect
may introduce systematic differences in the relative
Fe, Mg, and Na intensities.

The principal remaining uncertainty is the decomposition of
blended features, particularly the separation of Mg~I-2 from
neighbouring Fe emission around 517~nm. In the adopted model,
the relative strengths of individual transitions within each
component are constrained by atomic parameters and
temperature-dependent line emissivities, while the overall
Fe, Mg, and Na contributions are determined by three
independent scaling coefficients. This reduces the number
of freely adjustable line amplitudes but does not eliminate
uncertainties associated with spectral blending and the
adopted excitation temperature. The resulting component
intensities therefore remain model-dependent, particularly
when one component is weak relative to the surrounding emission.

Independent meteor velocities were not available for most events in the
present sample. The Fe--Mg--Na diagram is therefore used primarily as a
descriptive spectroscopic classification, without attempting a
velocity-dependent interpretation of the individual ternary positions.

The ternary positions should therefore be interpreted as a
spectroscopic description based on the relative integrated
intensities of the modelled Fe~I-15, Mg~I-2, and Na~I-1
components, rather than as measurements of elemental
abundances or bulk composition. In particular, displacement toward the Fe-rich
or iron region does not by itself demonstrate an iron meteoroid.
Conversely, the persistence of strongly Fe-dominated spectra after
explicitly accounting for the Fe contribution to the Mg region makes
such cases useful candidates for further physical interpretation when
independent trajectory, photometric, or other observational
information is available.

This distinction is relevant to the intended application of
spectroscopy within the Skytinel network. An independent example is
provided by the recent recovery of an iron meteorite observed by the
network. Spectral observations of that event showed dominant Fe/Ni
emission and suggested an iron meteoroid before the meteorite was
recovered \citep{Zmija2026Iron}. That event is not included in the
present sample, and its detailed spectroscopic analysis is beyond the
scope of this study and will be presented separately. It nevertheless
illustrates the potential value of spectroscopic information during
the analysis of potential meteorite falls.

Spectroscopy can therefore provide information complementary to
trajectory and photometric measurements. It does not determine bulk
density directly, but a spectroscopic indication of broad material
type may help constrain the range of physical properties considered
in subsequent modelling. Since bulk density is an important parameter
in dark-flight calculations, such an independent constraint may help
reduce the range of plausible solutions and consequently improve the
definition of a predicted meteorite search area.

\section{Conclusions}\label{sec:conclusions}

We developed a uniform procedure for comparing low-resolution meteor
spectra obtained at four spectroscopic stations of the Skytinel
network. After instrumental and atmospheric corrections, the relative
Fe~I-15, Mg~I-2, and Na~I-1 emission intensities were quantified
using a simplified spectral model based on selected atomic transitions
and temperature-dependent line emissivities. The resulting integrated
intensities were compared within the established Fe--Mg--Na
classification framework. The model explicitly accounts for blended
spectral features, including the contribution of Fe emission in the
vicinity of the Mg~I-2 triplet.

The analysed bright meteors show substantial spectral diversity. Most
of the spectra are compatible with the broad normal-meteor
distribution of \citet{Vojacek2015}, while individual events extend
toward the Fe-rich and Na-rich regions. SN48\_260521 represents the
most strongly Fe-dominated spectrum in the sample, with approximately
79\% of the summed modelled Fe~I-15, Mg~I-2, and Na~I-1 intensities
attributed to Fe~I-15. The results demonstrate that differences in
the relative strengths of the selected meteor spectral components
can be quantified even at the relatively low spectral resolution
of the Skytinel instruments.

The derived component ratios remain spectroscopic quantities and
should not be interpreted directly as elemental abundances, bulk
composition, or density. Their principal potential application is
therefore as an additional constraint used together with trajectory
and photometric information. In potential meteorite falls, an
independent spectroscopic indication of broad material type may help
restrict the range of physically plausible parameters adopted in
dark-flight calculations and thereby improve the definition of the
predicted search area.

\section*{ACKNOWLEDGMENTS}

The authors acknowledge the work of the Skytinel observers
who maintain the stations and collect meteor data.

NOIRLab IRAF is distributed by the Community Science
and Data Center at NSF NOIRLab, which is managed by
the Association of Universities for Research in
Astronomy (AURA) under a cooperative agreement with
the U.S. National Science Foundation.

\subsection*{Use of artificial intelligence tools}

OpenAI ChatGPT (GPT-based language models) was used
to assist with manuscript preparation, including
English-language editing, LaTeX formatting, and
the development and refinement of data-processing
scripts and mathematical descriptions of the
spectral model. All scientific assumptions,
analytical procedures, calculations, and results
were reviewed and validated by the authors,
who take full responsibility for the content
of this work.

\bibliography{Skytinel_meteor_spectroscopy_AN}

\section*{Supplementary Table S1. Spectral-model parameters}
\begin{table*}[htbp]
\setcounter{table}{0}
\renewcommand{\thetable}{S\arabic{table}}

\centering
\caption{Parameters used to reproduce the modelled Fe I, Mg I and Na I emission components for each meteor spectrum. The pseudo-Voigt profile uses Gaussian standard deviation $\sigma$ and Lorentzian half-width $\gamma$, both in nm. $C_{\mathrm{Fe}}$, $C_{\mathrm{Mg}}$ and $C_{\mathrm{Na}}$ are multiplicative coefficients in the line-emissivity model defined in the main text. }
\label{tab:supp_model_parameters}
\begin{tabular*}{\textwidth}{@{\extracolsep{\fill}}lrrrrrr@{}}
\hline
Spectrum & $T$ (K) & $\sigma$ (nm) & $\gamma$ (nm) & $C_{\mathrm{Fe}}$ & $C_{\mathrm{Mg}}$ & $C_{\mathrm{Na}}$ \\
\midrule
SN01-240813 & 4500 & 0.21213 & 0.80000 & 18 & 580 & 0.27 \\
SN01-250105 & 5000 & 0.21213 & 0.80000 & 15 & 150 & 0.08 \\
SN01-250120 & 5900 & 0.21213 & 0.80000 & 7 & 20 & 0.035 \\
SN38-260404 & 5900 & 0.21213 & 0.80000 & 3 & 12 & 0.06 \\
SN38-260912 & 5000 & 0.21213 & 0.80000 & 6 & 43 & 0.13 \\
SN47-260521 & 5900 & 0.21213 & 0.80000 & 3.5 & 15 & 0.045 \\
SN48-260103 & 5900 & 0.21213 & 0.80000 & 3 & 12 & 0.04 \\
SN48-260216 & 5900 & 0.21213 & 0.80000 & 2.7 & 12 & 0.015 \\
SN48-260521 & 7000 & 0.14142 & 0.80000 & 3.5 & 1 & 0.01 \\
SN48-260603 & 5700 & 0.21213 & 0.80000 & 2.9 & 18 & 0.04 \\
SN48-260626 & 5500 & 0.21213 & 0.80000 & 3.2 & 20 & 0.04 \\
SN48-260731 & 8000 & 0.21213 & 0.80000 & 1.3 & 0.7 & 0.04 \\
\bottomrule
\end{tabular*}
\end{table*}

\end{document}